\documentclass[]{iopart}

\usepackage{graphicx}
\begin{document}

\title[]{Dynamical percolation transition in two dimensional ANNNI model}

\author{Anjan Kumar Chandra}

\address{Theoretical Condensed Matter Physics Division, Saha Institute of Nuclear 
Physics, 1/AF Bidhannagar, Kolkata-700064, India}
\ead{anjanphys@gmail.com}
\begin{abstract}
The dynamical percolation transition of two dimensional Axial Next Nearest Neighbour Ising (ANNNI) model to pulsed magnetic field
has been studied by finite size scaling analysis (by Monte Carlo simulation)
for various values of frustration parameters, pulse width and temperature 
(below the corresponding static transition temperature).
It has been found that the size of the largest geometrical cluster shows a 
transition for a critical field amplitude. Although the transition points shift,
the critical exponents remain invariant for a wide range of frustration parameters.
It is also same as that obtained for the 2d Ising model. This suggests that
although the static phase diagrams of these two models differ significantly in 
various aspects, the dynamical percolation transition of both these models
belong to the same universality class. 

\end{abstract}


\maketitle

\section{Introduction}

The response of magnetic systems to time dependent external magnetic fields has
been of current interest in statistical physics
(\cite{Acharyya1}-\cite{Korniss2}).
Magnetisation reversal dynamics is an important subject for many technological 
applications of magnetic materials. In magnetic recording industry fast
magnetisation reversal in storage media is essential for high data transfer rate 
\cite{Vogel}. These spin systems driven by time-dependent external magnetic 
fields, have 
basically got a competition between two time scales: the time period of the 
driving field and the relaxation time of the driven system, giving rise to 
interesting non-equilibrium phenomena. Magnetisation reversal by pulsed magnetic 
field has been studied extensively (\cite{Acharyya1}-\cite{Korniss2}). 
The magnetisation reversal were studied below the corresponding static
transition temperature ($T_c^0$), where the system remains ordered. A pulse
of finite strength ($h_p$) and finite duration ($\Delta t$), applied along
the opposite direction of the prevalent order tends to reverse the magnetisation.
Depending on the temperature ($T$) and pulse duration ($\Delta t$), there
will be a critical field strength ($h_p^c$) at which the system will be forced
to change from one ordered state (magnetisation $m_0$) to another ordered state
(magnetisation $-m_0$). Below this critical strength, the magnetisation will
eventually remain unchanged as the field pulse is taken off. This transition
is called ``magnetisation-reversal transition".

A percolation transition is also
associated with this magnetisation reversal induced by the external magnetic
field. By percolation transition we mean the transition of the largest 
``geometrical cluster" (consisting of nearest neighbor parallel spins)
size from a large value to zero. In the initially ordered state, large clusters
of particular spin orientation (``up" or ``down") exists. 
But due to the pulsed field
directed along the opposite direction, the spins gradually flip (depending on the magnitude and duration) and for a critical value of the field or pulse duration
the pre-existing large clusters vanish, giving rise to oppositely oriented 
clusters. Thus if we measure the largest ``geometrical cluster" size for a 
particular 
orientation at a certain temperature (below the critical temperature) and finite 
duration of field pulse, it shows
a transition at a critical field magnitude. Although thermal percolation 
transition of the ``geometrical" clusters,
in 2d Ising model has been studied extensively \cite{muler1,stoll,muler2,binder1},
very little is known about the field induced dynamical percolation transition
\cite{Biswas}. 
It was found numerically (based on Monte Carlo simulation) that 
for pure two dimensional Ising model
at finite temperature the field induced magnetisation transition and the 
percolation transition occur for the same critical magnetic field. 
Moreover the critical exponents were found (using finite size scaling analysis)
different from that of the thermal percolation transition, indicating the
dynamical percolation transition to belong to a separate universality class 
\cite{Biswas}. This transition was also studied for the 2d Ising model with
diagonal next nearest neighbor interaction for a typical value of the frustration
parameter \cite{Biswas} and it was found that the critical exponents remain 
unchanged.

\begin{center}
\begin{figure}
\includegraphics[width=3in]{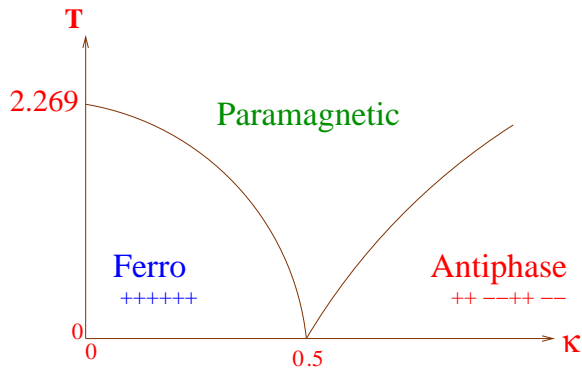}
\caption{\label{fig:ani}Schematic phase diagram of the two-dimensional ANNNI
model according recent studies \cite{Shirahata,Chandra}
}
\end{figure}
\end{center}

Here we intend to study the field induced percolation transition for the
2d Axial Next Nearest Neighbour Ising (ANNNI) model which is one of the simplest 
classical Ising model with a tunable frustration and has been studied for a 
long time (see \cite{selke1,selke2,yeomans,liebmann} for review). The 2d ANNNI 
model is a square lattice Ising model with 
nearest-neighbor ferromagnetic interaction along both the axial directions
and second-neighbor anti-ferromagnetic interaction along one axial direction.
The Hamiltonian is
\begin{equation}
{\mathcal H} = - J \sum_{x,y} s_{x,y} [s_{x+1,y}+s_{x,y+1}-\kappa s_{x+2,y}]
\end{equation}
where the sites ($x$,$y$) run over a square lattice, the spins $s_{x,y}$ 
have binary states
($\pm 1$), $J$ is the nearest-neighbor interaction strength and $\kappa$ is a
parameter of the model. For positive values of $\kappa$ the second-neighbor
interaction introduces a competition or frustration. The $T-\kappa$ phase 
diagram for this
model shows a rich behavior. It is easy to prove analytically that
\cite{selke1,selke2,yeomans,liebmann} at
zero temperature, the system is in a ferromagnetic state for $\kappa < 0.5$,
and in antiphase ($++--++-- \cdots$ along $x$ direction and all like spins
along $y$ direction) for $\kappa > 0.5$ with a ``multiphase'' state at
$\kappa = 0.5$. (The multiphase state comprises of all possible configurations
that have no domain of length 1 along the $x$ direction.) But the finite
temperature phase diagram is yet to be solved analytically. There had been
many predictions from various kinds of numerical calculations. But here we
present the qualitative phase diagram (Fig.~\ref{fig:ani}) that has been 
obtained by extensive Monte Carlo studies \cite{Shirahata,Chandra}.
As the previously conjectured floating phase was not found in \cite{Shirahata,Chandra},
it has not been shown in Fig.~\ref{fig:ani}.

The phase diagram consists of ferromagnetic phase 
(for $\kappa < 0.5$) and 
antiphase (for $\kappa > 0.5$) at low temperature along with a paramagnetic 
phase at high temperature (for all $\kappa$ values). So for both $\kappa < 0.5$
and $\kappa > 0.5$ there is a transition temperature depending upon the value
of $\kappa$. For $\kappa < 0.5$, the study of percolation transition is
done in a similar way as that for the unfrustrated 2d Ising model. But
for $\kappa > 0.5$, the cluster cannot be defined as we have for $\kappa < 0.5$.
Because for $\kappa > 0.5$ below the static transition temperature we mainly have
striped states with like spins oriented along $Y$-axis
and a periodic modulation along the direction of frustration.
Here we propose an alternative definition of cluster to study the percolation
transition. By a cluster we mean a set of nearest neighbour spins which is in the
same state as it was in its initial condition. Starting from an exactly 
antiphase state
(initial cluster size is the system size) we acheive a steady state at a 
particular temperature (due to thermal
agitation some of the spins flip and thus patches of clusters form). Now 
we apply a field pulse which varies spatially along the system such that
the field acts on the spin along the direction opposite to that with which
it started the dynamics i.e. $sgn(h_p)=sgn(s_{x,y}(0))$, where $h_p$ is the
applied field strength at site $x,y$ and $s_{x,y}(0)$ is the initial spin 
value at lattice site $x,y$. This means that if we start with a 
``$++--++-- \cdots$" spin configuration
along $x$ direction, the applied field will be along downward direction for
the first two columns and upward along the next two columns and this pattern 
continues over the entire system.  
As soon as we switch on the field pulse (applied along the reverse direction of 
the initial spin orientation), the spins will tend to orient along the field
direction and the spins which are at a state identical to that of its initial 
state
may flip (depending on the field strength and duration) and thus the cluster 
sizes eventually decreases giving rise to a percolation transition. 

In this paper we study the percolation transition for a wide range of
frustration parameter ($0.0<\kappa<1.0$) and find how the critical
field strength varies with the frustration parameter for a particular 
pulse duration. We have used Monte Carlo simulation to obtain the numerical
results and fully periodic boundary condition throught our entire study. We have 
calculated the critical exponents numerically
by finite size scaling analysis to see whether the exponents are affected
by the frustration parameter. In Section $2$ we discuss the model studied.
In Section $3$, we present the results obtained for the finite size
scaling analysis of the percolation order parameter and in Section $4$
the fractal dimension and the scaling analysis of the Binder cumulant.
In the final section we have discussed some consequences related to our
study. The main observation of this work is that the critical exponents
are independent of the frustration parameter and also same as that of
the normal $2d$ Ising model.

\section{The Model}

The model studied here is the 2d ANNNI model under a time-dependent external
magnetic field pulse, described by the Hamiltonian
\begin{equation}
{\mathcal H} = - J \sum_{x,y} s_{x,y} [s_{x+1,y}+s_{x,y+1}-\kappa s_{x+2,y}]
- h(t) \sum_{x,y} s_{x,y}
\end{equation}
where the sites ($x$,$y$) runs over a square lattice, the spins $s_{x,y}$ are
$\pm 1$, $J$ is the nearest-neighbor interaction strength (we keep $J=1.0$ 
here on) and $\kappa$ is the competition or frustration parameter. 
The static critical temperature ($T_c^0$) depends on 
the value of $\kappa$ (Fig.~\ref{fig:ani}). For $\kappa = 0$, 
$T_c^0 = 2/\ln(1+\sqrt{2})\approx2.269...$. For $\kappa<0.5$, the field pulse 
applied is spatially uniform (directed opposite to that of the initial spin 
configuration we start with) having a time dependence as :\\
\begin{equation}
h(t) = \left\{
 \begin{array}{rl}
  -h_p & , t_0 \le t \le t_0+\Delta t\\
   0 & \mbox{otherwise}
\label{field-pulse}
 \end{array} \right.
\end{equation}
For $\kappa>0.5$, the field pulse applied is such that the field 
direction
is opposite to that of the starting configuration (antiphase order) of the 
spins, but the time dependence is the same as that of Eq. \ref{field-pulse}.
In our simulation we ensure that the time $t_0$ at which the pulse is 
applied,
the system reaches its equilibrium configuration at that temperature 
$T~ (< T_c)$. 

\begin{center}
\begin{figure}
\includegraphics[width= 6cm, angle = 270]{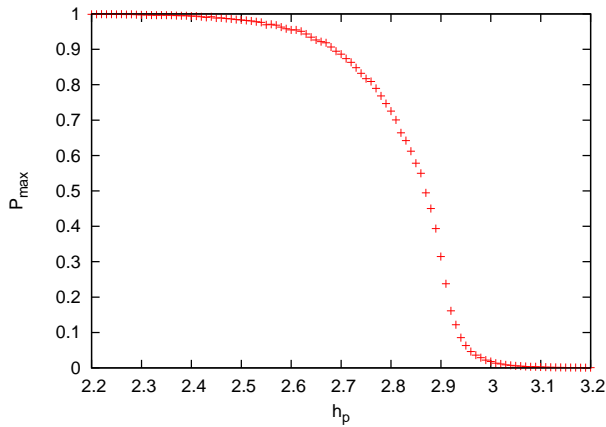}
\caption{\label{compare}Variations of the largest cluster size ($P_{max}$)
with the field pulse amplitude ($h_p$) in case of 2d ANNNI model
where $\kappa = 0.3$, $T=0.2$ and $\Delta t = 4$. 
}
\end{figure}
\end{center}

\begin{center}
\begin{figure}
\noindent \includegraphics[width= 6cm, angle = 270]{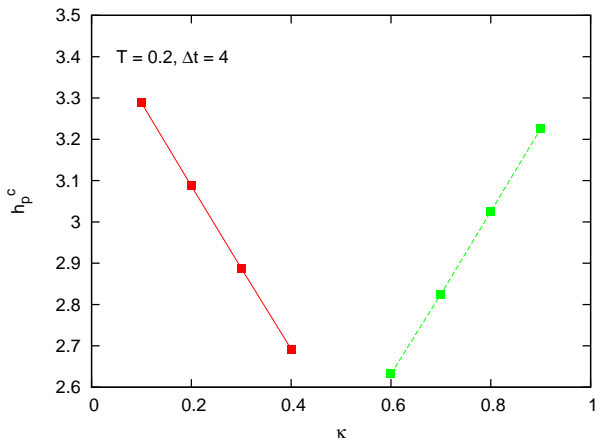}
\caption{\label{fig:kappa_hp} The plot of $h_p^c$ against the frustration 
parameter $\kappa$. (We have ommitted the study close to $\kappa=0.5$, because
in this region the transition temperatures are too low.)
}
\end{figure}
\end{center}

The spins are selected randomly and are flipped following the normal
Metropolis algorithm. The energy difference ($\Delta E$) to flip the spin 
is calculated and flipped with probability min\{$1,\exp(-\Delta E/T)$\}.
At $t=t_0$, the system acheives a steady state at the corresponding 
temperature (in the absence of any field). As soon as the field pulse
is switched on, a large number of spins which were intact in their initial state, 
flip during the pulse duration.
Thus the percolation order parameter $P_{max} = S_L/L^2$ (where $S_L$
is the size of the largest cluster and $L$ is the linear size of the
system) decreases during this duration and for a particular combination
of $T, \kappa$ and $\Delta t$, for a particular value of $h_p^c$, the
system undergoes a percolation transition (Fig. \ref{compare}).

It is found that for a particular temperature for $\kappa<0.5$, as the value of 
$\kappa$ is increased, the
transition field $h_p^c$ decreases, but for $\kappa>0.5$, $h_p^c$ increases
with $\kappa$ (Fig.~\ref{fig:kappa_hp}). For $\kappa<0.5$, as $\kappa$ is
increased, the increasing frustration itself destabilises the order and thus 
the pulse amplitude needed for transition decreases. But for 
$\kappa>0.5$, the situation is different. The increasing frustration parameter
stabilises the antiphase more and more, thus requiring higher pulse amplitude for
the percolation transition (Fig.~\ref{fig:kappa_hp}).

\section{Finite size scaling analysis of the percolation order parameter}

\begin{figure}
\noindent \includegraphics[width= 8cm]{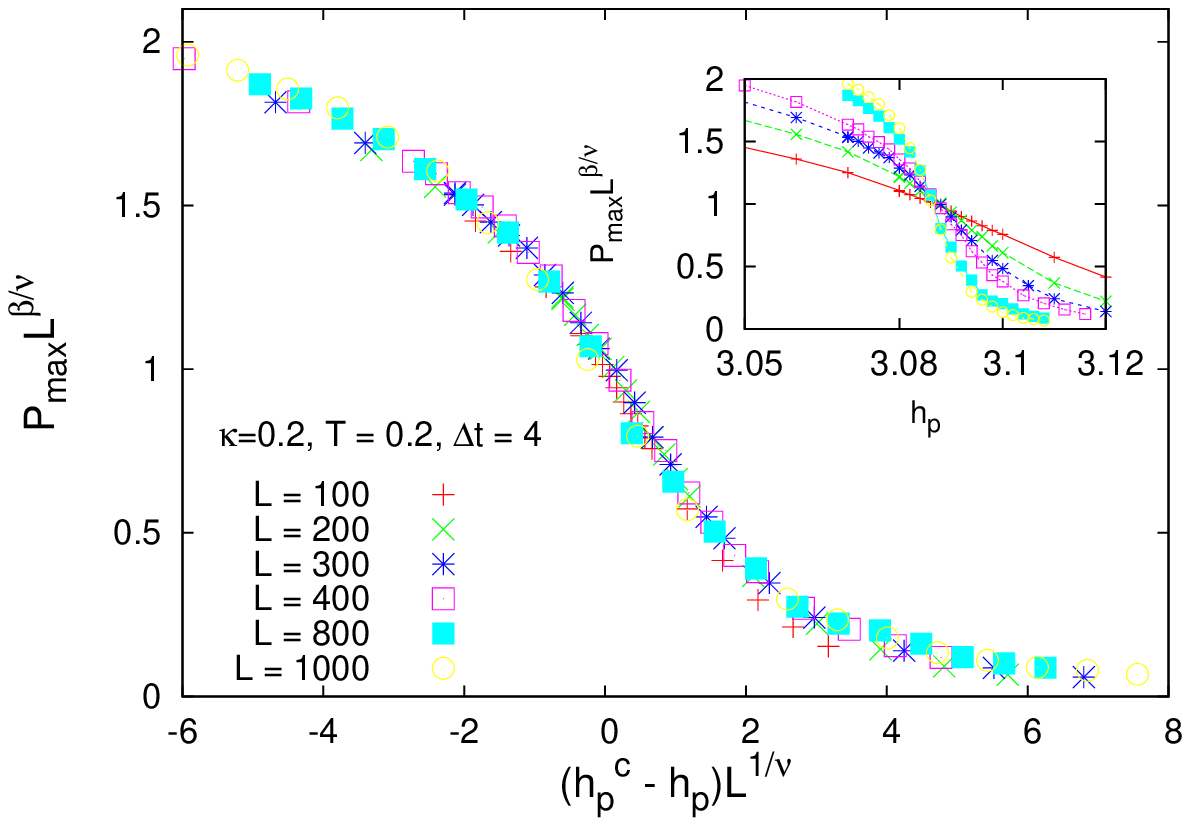}
\caption{\label{fig:perc-dist4.klt} Inset : The scaled largest cluster size 
($P_{max}L^{\beta/\nu}$) has been plotted with $h_p$ for six different system sizes 
($L = 100, 200, 300, 400, 800$ and $1000$) for a typical value of $\kappa<0.5$. The 
crossing point determines the critical field amplitude ($h_p^c$). Main : The value
of $h_p^c$ and $\beta/\nu$ has been used to make the plots for the scaled 
variables
$P_{max}L^{\beta/\nu}$ and $(h_p^c - h_p)L^{1/\nu}$ collapse by tuning the value 
of $1/\nu$ for all the different $L$ values.
}
\end{figure}

\begin{figure}
\noindent \includegraphics[width= 8cm]{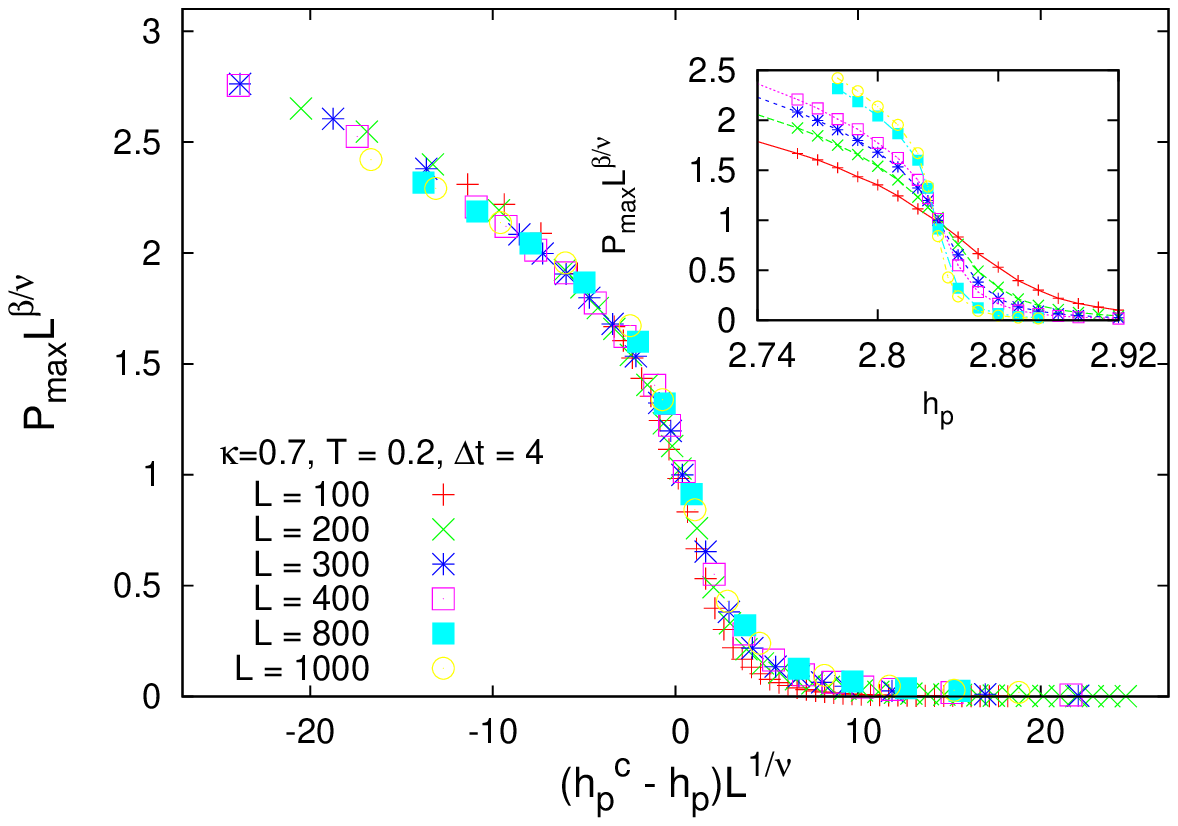}
\caption{\label{fig:perc-dist6.kgt} Inset : The scaled largest cluster size 
($P_{max}L^{\beta/\nu}$) has been plotted with $h_p$ for six different system sizes 
($L = 100, 200, 300, 400, 800$ and $1000$) for a typical value of $\kappa>0.5$. The 
crossing point determines the critical field amplitude ($h_p^c$). Main : The value
of $h_p^c$ and $\beta/\nu$ has been used to make the plots for the scaled variables
$P_{max}L^{\beta/\nu}$ and $(h_p^c - h_p)L^{1/\nu}$ collapse by tuning the value 
of $1/\nu$ for all the different $L$ values..
}
\end{figure}

In this section we discuss about the percolation transition behaviour.
For a particular combination of $\kappa, T$ and $\Delta t$, we study the 
variation of $P_{max}$ with $h_p$. The percolation transition is characterised
by several exponents which also determine the universality class. The order parameter 
i.e., the relative size of the largest
cluster varies as,
\begin{equation}
P_{max}\sim (h_p^c-h_p)^{\beta}.
\end{equation}

\begin{figure}
\noindent \includegraphics[width= 6cm, angle=270]{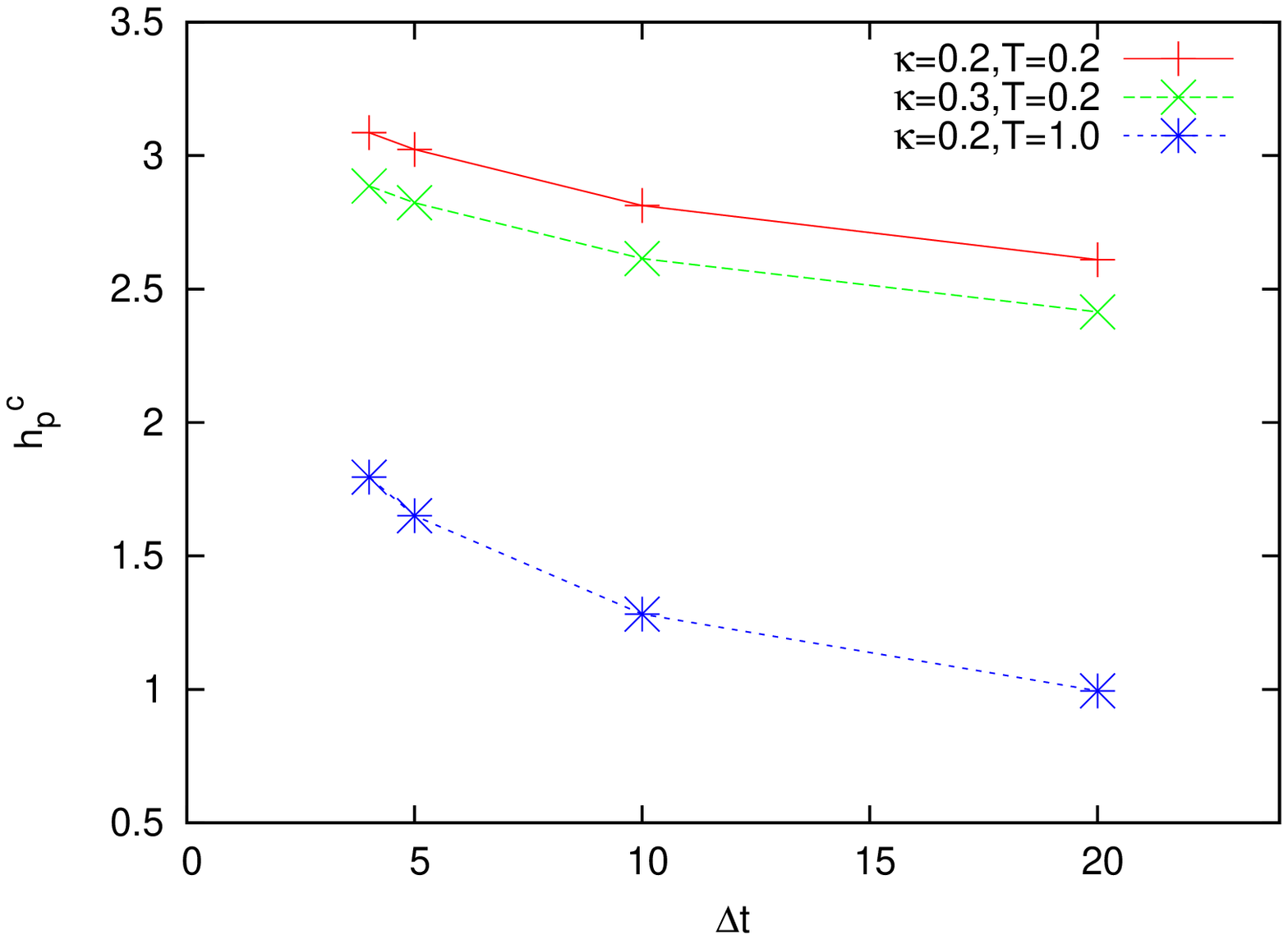}
\caption{\label{fig:hp_deltat_klt5} The plot of $h_p^c$ against $\Delta t$ for
three different combinations of ($\kappa,T$) for the regime $\kappa<0.5$.
}
\end{figure}

\begin{figure}
\noindent \includegraphics[width= 6cm, angle=270]{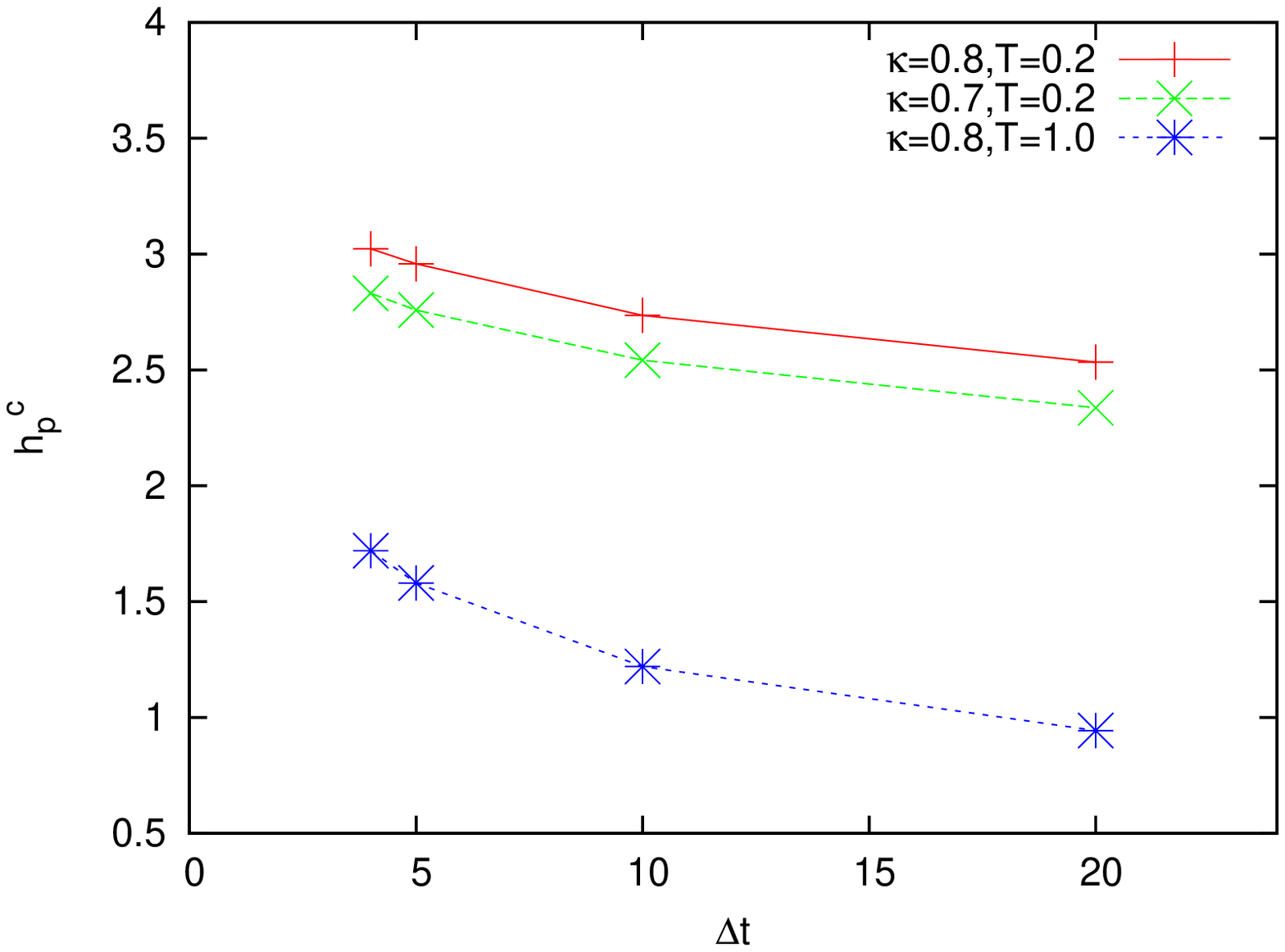}
\caption{\label{fig:hp_deltat_kgt5} The plot of $h_p^c$ against $\Delta t$ for
three different combinations of ($\kappa,T$) for the regime $\kappa>0.5$.
}
\end{figure}

\noindent The correlation length diverges near the percolation transition point as
\begin{equation}
\xi \sim (h_p^c-h_p)^{-\nu},
\end{equation}
\noindent where, $h_p^c$ is the critical field amplitude.
The other exponents can be calculated from various scaling relations \cite{Stauffer}.
But due to finite-size effects it is difficult to determine these exponents precisely. 
We have determined these exponents from the finite size scaling relations.  
$P_{max}$ follows the scaling form
\begin{equation}
P_{max}=L^{-\beta/\nu}\mathcal{F}\left[L^{1/\nu}\left (h_p^c-h_p\right)\right],
\end{equation}
where $\mathcal{F}$ is a suitable scaling function. By tuning $\beta/\nu$, all the 
$P_{max}L^{\beta/\nu}$-$h_p$ curves can be made to cross at a single point and
this point determines the critical field amplitude ($h_p^c$). Now once $\beta/\nu$
and $h_p^c$ has been estimated, by tuning $1/\nu$ one can collapse all the 
$P_{max}L^{\beta/\nu}$-$(h_p^c-h_p)L^{1/\nu}$ curves on one another, and thus
$\nu$ is determined. We have shown both these plots for $\kappa=0.2$ ($\kappa<0.5$)
and $\kappa=0.7$ ($\kappa>0.5$) in Fig.~\ref{fig:perc-dist4.klt} and 
Fig.~\ref{fig:perc-dist6.kgt} respectively. For both the curves we get the
same values for $\beta/\nu = 0.20\pm0.05$ and $1/\nu = 0.85\pm0.05$ and more
importantly this value is the same as that we obtained for the normal 2d Ising model
found earlier \cite{Biswas} which indicates that the dynamical percolation 
transition of the 2d ANNNI model also belongs to the same universality class
as that of the normal 2d Ising model. 
Only the critical field amplitude changes (as shown in Fig.~\ref{fig:kappa_hp}). 
We have repeated the same exercise for some other combination of $\kappa, T$ and 
$\Delta t$, but the exponents remain unchanged. This suggests that the
magnitude of frustration does not have any drastic effect on the dynamical
percolation phenomenon except the quantitative shift of the transition point.

To have a picture of the effect of various parameters on the critical field amplitude
we have given two graphs, one for $\kappa<0.5$ (Fig.~\ref{fig:hp_deltat_klt5}) and the 
other for $\kappa>0.5$ (Fig.~\ref{fig:hp_deltat_kgt5}), each 
showing the variation of $h_p^c$ with $\Delta t$ for different sets of $\kappa$
and $T$. It is quite clear from Fig.~\ref{fig:hp_deltat_klt5} that, for any
combination of $\kappa$ and $T$, $h_p^c$ decreases with $\Delta t$. Also for
any comination of $\kappa$ and $\Delta t$, $h_p^c$ decreases with $T$, because
thermal excitation always helps in percolation transition and thus reducing
the necessary external field amplitude. But the frustration parameter plays
an opposite role on the two halves of the $\kappa$ regime. For a fixed pair of
$T$ and $\Delta t$, when we are below $\kappa<0.5$, with increase of $\kappa$ we need 
less field for reaching the transition point because increasing the frustration
parameter itself reduces the parallel alignment between the adjacent spins and
thus facilitating the transition. But above $\kappa=0.5$, the increase in frustration
parameter supports the antiphase structure (with which we begin the dynamics for
$\kappa > 0.5$), and thus we have to apply more field amplitude to flip the 
spins to opposite direction. Thus the frustration parameter plays opposite role
in cluster formation in the two regimes. 

\section{Fractal dimension and Binder cumulant}

\begin{figure}
\noindent \includegraphics[width= 6cm, angle=270]{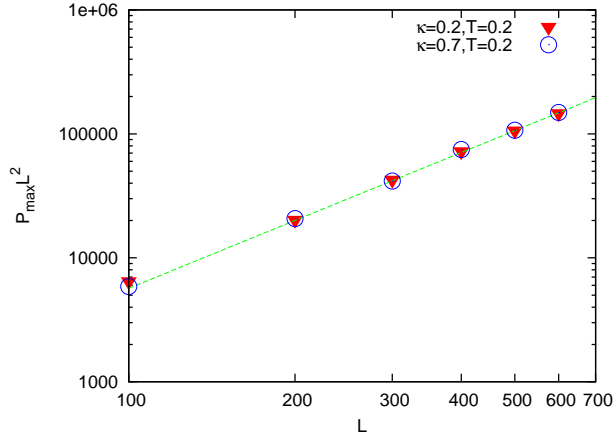}
\caption{\label{fig:fractald} Variation of the largest cluster size with the
linear system size for two different values of $\kappa = 0.2$ and $0.7$ at the
critical field for $T = 0.2$ and $\Delta t = 4$. The log-log plot gives the
fractal dimension to be $D=1.82\pm0.01$ for both the values of $\kappa$.
}
\end{figure}

At the transition point, the percolation clusters become self similar. The
largest cluster depends on the linear system size and it varies as $S_L \sim L^D$
where $S_L$ is the largest cluster and $D$ is the fractal dimension.
The fractal dimension measured (Fig.~\ref{fig:fractald}) is found to be 
independent of the value of $\kappa$ and the value is $D = 1.82\pm0.01$ which is
again the same as that we found in case of the normal 2d Ising model. This
fractal dimension is known to be related to the spatial dimension ($d$) as 
$D = d - \beta/\nu$. This relation is also satisfied for our estimated values
of $\beta/\nu = 0.20\pm0.05$ (mentioned in previous section) and $D$ for
all values of $\kappa$. 

\begin{figure}
\noindent \includegraphics[width= 8cm]{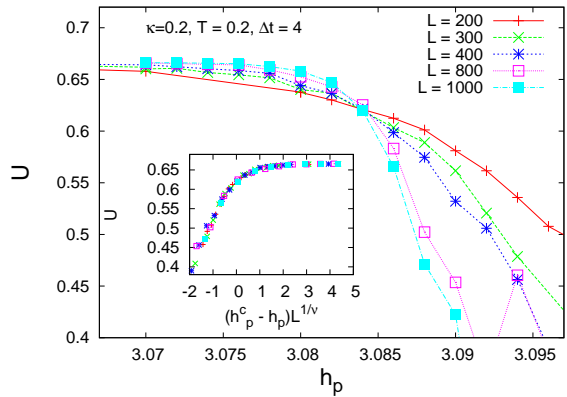}
\caption{\label{fig:perc-dist4.klt.binder} Main : The Binder cumulant has been 
plotted with $h_p$ for six different system sizes
($L = 200, 300, 400, 800$ and $1000$) for a typical value of $\kappa<0.5$. The
crossing point determines the critical field amplitude ($h_p^c$). Inset : The value
of $h_p^c$ has been used to make the plots for $U$ and $(h_p^c - h_p)L^{1/\nu}$ 
collapse by tuning the value of $1/\nu$ for all the different $L$ values.
}
\end{figure}

\begin{figure}
\noindent \includegraphics[width= 8cm]{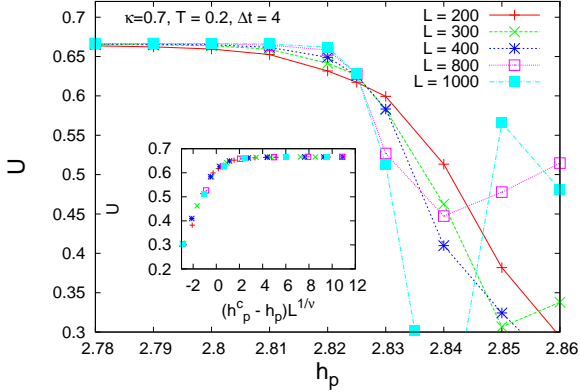}
\caption{\label{fig:perc-dist6.kgt.binder} Main : The Binder cumulant has been        
plotted with $h_p$ for six different system sizes
($L = 200, 300, 400, 800$ and $1000$) for a typical value of $\kappa>0.5$. The
crossing point determines the critical field amplitude ($h_p^c$). Inset : The obtained value
of $h_p^c$ has been used to make the plots for $U$ and $(h_p^c - h_p)L^{1/\nu}$
collapse by tuning the value of $1/\nu$ for all the different $L$ values.
}
\end{figure}

For further verification of the critical points and the exponents, we have also
studied the Binder cumulant of the order parameter, defined as \cite{Binder},
\begin{equation}
U=1-\frac{\langle P_{max}^4\rangle}{3\langle P_{max}^2\rangle^2},
\end{equation}
where $P_{max}$ is the percolation order parameter and the angular brackets
denote the ensemble average. The crossing point of the curves $U-h_p$ for different
system sizes gives the critical field amplitude ($h_p^c$). The Binder cumulant has 
also been used to
determine the correlation length exponent as it follows the scaling form
\begin{equation}
\label{binder-nu}
U=\mathcal{U}((h_p^c-h_p)L^{1/\nu}),
\end{equation}
where $\mathcal{U}$ is a suitable scaling function. Having obtained $h_p^c$ from the 
crossing point, if we plot $U$ with the scaled 
field amplitude $(h_p^c-h_p)L^{1/\nu}$, for a suitable value of
$\nu$ we will obtain a data collapse and that value of $\nu$ determines 
the correlation length exponent.
We have studied these results both 
for $\kappa<0.5$ (Fig.~\ref{fig:perc-dist4.klt.binder}) and $\kappa>0.5$ 
(Fig.~\ref{fig:perc-dist6.kgt.binder}).  

For all values of $\kappa$ (only two sets of $\kappa,T$ has been presented in 
Fig.~\ref{fig:perc-dist4.klt.binder} and Fig.~\ref{fig:perc-dist6.kgt.binder}), the 
value of $1/\nu$ remains same ($0.85\pm0.05$) and also equal to that we have already
obtained from that of the finite size scaling of percolation order parameter.
The value of the Binder cumulant at the crossing point for various system sizes is 
also identical ($0.62\pm0.01$). 
The only obvious quantitative change is seen for the transition point.

\section{Discussions}

Spatially modulated phases of metallic alloys can be investigated in terms of 
ANNNI model and considering the application of these alloys in magnetisation
reversal phenomenon, we have studied the magnetic field-pulse induced
percolation phenomenon in 2d ANNNI model.
We have focused on the role of the frustration parameter in affecting the
percolation transition of the 2d ANNNI model.
By applying magnetic field-pulse of finite duration ($\Delta t$)
for a certain frustration parameter ($\kappa$) and temperature ($T$) (below
the corresponding static transition temperature), we have determined 
the critical field strength ($h_p^c$) for which the transition occurs. 
Although we have
found that the frustration parameter highly enriches the static phase diagram
of 2d ANNNI model (various novel phases and transitions appear due to $\kappa$),
quite surprisingly it does not have any large impact on the dynamical percolation
transition apart from shifting the transition points. We have restricted our
study in the regime $0.0<\kappa<1.0$. From finite size scaling analysis we have 
showed that 
both the percolation order parameter and the Binder cumulant leads to the   
same values of critical exponents as we have found for the normal 2d Ising model
(thus it belongs to the same universality class as that of the dynamical percolation
transition of 2d Ising model).
The values of exponents determined are $\beta/\nu = 0.20\pm0.05$ and
$1/\nu = 0.85\pm0.05$. 
The fractal dimension $D = 1.82\pm0.01$ also confirms the relations between the 
exponents.

Although we have not presented the details, in this context
we have also investigated the transition behaviour of the magnetisation (for 
$\kappa<0.5$) and sub-lattice magnetisation (for $\kappa>0.5$). 
The sub-lattice magnetisation is defined as,
$x$-direction, the sub-lattice magnetization is defined by
$m_{s} = \frac{1}{L} \sum^{\frac{L}{4}-1}_{x=0} (m_{4x+1} + m_{4x+2} -
m_{4x+3} - m_{4x+4})$ 
with,
$m_x = \frac{1}{L} \sum^L_{y=1}  s_{x,y} $
where $L$ is the length of the system.
It has been 
found that the transition points were same as that of the percolation
transition (for a certain set of value of $\kappa, T$ and $\Delta t$).
The study of finite-size scaling of the order parameters also gives the same 
critical exponents ($\beta/\nu$ and $1/\nu$). Only the value of the crossing 
point of the Binder cumulant for magnetisation shows non-universal behaviour
for different temperatures. This value ($U^{\star} = 0.62\pm0.01$) is robust in 
case of percolation transition.

This study may be useful in the application of magnetisation reversal of spatially
modulated magnetic materials where the only change necessary is the value
of the magnitude of the critical field, keeping the nature of the transition
intact. Although we have limited our regime of investigation below $\kappa=1.0$,
one can also explore this transition beyond $\kappa=1.0$ or other modulated
structures. 

\section{Acknowledgements}
The author acknowledges many fruitful discussions and suggestions of
Prof. Bikas K. Chakrabarti, Prof. Subinay Dasgupta and Mr. Soumyajyoti Biswas. The 
author acknowledges the financial support from
DST (India) under the SERC Fast Track Scheme for Young
Scientists Sanc. No. SR/FTP/PS-090/2010(G).

\end{document}